\documentstyle{article}

\begin{document}

\title{Bosonic Super Liouville System: Lax Pair and Solution}

\vspace{2cm}

\author{{Liu Zhao$^{a,b}$   and   Changzheng Qu$^{a,c}$}\\
{\small $^{a}$CCAST (World Lab), Academia Sinica, 
P O Box 8730, Beijing 100080, P R China}\\
{\small $^{b}$Institute of Modern Physics, 
Northwest University, Xian 710069, P R China}\\
{\small $^{c}$Department of Mathematics, 
Northwest University, Xian 710069, P R China}}

\date{December 3, 1996}

\maketitle

\thispagestyle{empty}

\begin{center}

\begin{minipage}{120mm}

\vskip 0.5in

{We study the bosonic super Liouville system which is a statistical
transmutation of super Liouville system. Lax pair for the bosonic 
super Liouville system is constructed using prolongation method, 
ensuring the Lax integrability, and the solution to the equations 
of motion is also considered via Leznov-Saveliev analysis.}

\vspace{2cm}


\end{minipage}
\end{center}

\vskip 0.5in

\newpage

\section{INTRODUCTION}

Liouville and super Liouville equations are found to be important
in quite a vast range of physical problems. For examples, Liouville 
equation is closely connected to string theory, two-dimensional
gravity in the conformal gauge and is a very popular example of
two-dimensional integrable field theory with conformal invariance,
and the same roles in super analogs of the above problems are played
by super Liouville equation. 

From the point of view of Toda lattice field theory, the Liouville 
equation is nothing but the simplest Toda field theory with
the Toda lattice denoted by a single node--the Dynkin diagram of the 
Lie algebra $sl(2)$ (and the super Liouville equation, 
which gauges the basic Lie superalgebra $osp(1|2)$ (Toppan 1991), 
is the simplest one from the family of super Toda field theories).
It is remarkable that for each underlying 
Lie algebra ${\cal G}$ 
one can construct a Toda field theory. Analogously, for each 
basic Lie superalgebra 
%\footnote{It is recently shown that super Toda theories
%can be established for non basic Lie super algebras also (Toppan 1996)}
one can construct a super Toda field theory.
A more interesting fact is that for each Lie algebra ${\cal G}$ of
rank $r >1$ there exists a so-called bosonic super Toda theory
(Chao (1993), Chao \& Hou (1993, 1994), Hou \& Chao (1993)), 
a kind of lattice field theory which can be viewed as the usual Toda 
field theory coupled to some bosonic ``matter'' fields, 
whose equations of motion
looks very similar to the super Toda equations written in component form
except the following two points: 
\footnote{Compare the equations of motion for bosonic super Toda 
theories in Chao (1993), Chao \& Hou (1993, 1994), Hou \& Chao (1993)
and that of supersymmetric Toda theories in, e.g., Au \& Spence (1995).}
i) the Cartan matrices entering the 
equations of motion are different for bosonic super and true super Toda 
theories since the underlying algebras are different; 
ii) the bosonic super Toda theory contains only bosonic
fields and hence does not yield a true supersymmetry. However, despite these 
differences, the bosonic super Toda theory does yield very nice
mathematical properties both as integrable and conformal field theoretic
models, in particular, such a model is intimately related to the 
$W_{n}^{(2)}$ algebra if the underlying gauge group is chosen to be 
$SL(n,~R)$ (Chao \& Hou (1994)). Moreover, it was recently argued 
in Ferreria {\it et al} (1995) and Gervais \& Saveliev (1995) that 
though classically
the extended Toda theories such as the bosonic super Liouville theory
contains only bosonic fields, their quantum versions might give rise some
fermionic degrees of freedom and may have relevant applications in
photo-electronic problems. Both due to the mathematical beauty and the 
potential physical significance, much efforts have been paid to the study of 
bosonic super Toda theories (Chao (1993, 1994, 1995), Chao \& 
Hou (1993, 1994), Hou \& Chao (1993)).

Two puzzling points are worth of further efforts in the above picture. First,
since the equations of motion for bosonic super Toda and true super
Toda theories are so much alike, one naturally expects to establish
some relationship between these two kinds of theories; Second, though 
the bosonic super Toda theory exists for almost all underlying Lie
algebras, the simplest rank one Lie algebra $sl(2)$ is excluded from 
this picture and therefore no bosonic super Liouville model exists
along the above line.

A naive answer to the first problem might be such that the bosonic 
super Toda theories are just the statistically transmuted super Toda 
theories, i.e. by replacing all the fermionic fields in super Toda 
theories by bosonic ones one get a bosonic super Toda theory. But this 
cannot be true as is mentioned, the Cartan matrices entering the equations 
of motion are quite different for both kinds of theories. However, this 
naive idea might be a useful clue to construct a bosonic super Liouville
model and in this paper we do adopt such a technique to define a
bosonic super Liouville model.

\indent We start from the supersymmetric Liouville equation

\begin{equation}
D_{+}D_{-}\Phi={\rm exp}(\Phi), \label{sl}
\end{equation}

\noindent where we have chosen
$D_\pm =\frac{1}{3}\frac{\partial}{\partial \theta_\mp} \pm 
\theta_\mp \frac{\partial}{\partial x_\pm}$ and

\begin{displaymath}
\Phi= \phi + 3\sqrt{2}\left(\theta_+ \psi_- + \theta_- \psi_+ \right)
+ 6 \theta_+ \theta_- F
\end{displaymath}

\noindent so that, in component form, equation (\ref{sl}) can be 
rewritten as follows

\begin{eqnarray}
& &\partial_{+}\partial_{-}\phi=18\psi_{+}\psi_{-}e^{\phi}+4e^{2\phi}, 
\nonumber\\
& &{\partial_{+}\psi_{-}}=3\psi_{+}e^{\phi}, \nonumber\\
& &\partial_{-}\psi_{+}=3\psi_{-}e^{\phi}.    \label{comp}
\end{eqnarray}

\noindent We see that these equations have exactly the same form
as some extended Liouville equation obtained by one of the author earlier
in Chao (1993, 1994) except that the fields $\psi_\pm$ 
in (\ref{comp}) are fermionic.
We call equation (\ref{comp}) with $\psi_\pm$
changed into bosonic fields a statistically transmuted 
super Liouville equation or bosonic super Liouville equation (BSLE), and 
the present paper is just devoted to study the integrability of that 
equation. We stress that, in BSLE, no signature change occur in front of the 
$\psi_+ \psi_-$ term while the order of $\psi_+$ and $\psi_-$ is reversed.

Before going into detailed studies, let us mention that, eq. (\ref{comp}),
viewed as BSLE, represent the usual Liouville system coupled to 
a pair of external fields $\psi_\pm$. Moreover these external fields do not
possess mass, because the whole system of equations of motion is 
conformally invariant, i.e. if the coordinate system $(x_+,~x_-)$
undergo the following conformal transformation

\begin{displaymath}
x_\pm \rightarrow f_\pm(x_\pm),
\end{displaymath}

\noindent the equations of motion will be left invariant provided the fields
$\phi,~\psi_pm$ transform as

\begin{eqnarray}
\left\{ 
\begin{array}{l}
$$\phi \rightarrow + { \rm ln} (f_+')^{1/2} (f_-')^{1/2},$$\cr
$$\psi_\pm \rightarrow (f_\pm')^{1/2} \psi_\pm,
\end{array}
\right. \nonumber
\end{eqnarray}

\noindent It is interesting to see that the statistical transmutation
from the super Liouville equation to BSLE also changes the fields
$\psi_+$ and $\psi_-$ from the $(\frac{1}{2},~0)$ and $(0,~\frac{1}{2})$
of Lorentz group to that of the classical conformal group.

\section{LAX-PAIR AND SYMMETRY ALGEBRAS FOR BSLE}

In this section we shall address the problem of integrability
of the BSLE (\ref{comp}). A system of nonlinear partial differential 
equations is said to be integrabe if it is a Hamiltonian system and possesses 
an infinite number of Poisson commuting integrals of motion. This is the 
classical Liouville sense of integrability. Another slightly
weaker definition of integrability is that if the system can be 
identified to the compatibility condition of a system of 
linear auxiliary problems, i.e. the Lax pair. The Lax integrability 
will be identical to Liouville integrability if the Lax system 
admits a fundamental Poisson structure and this Poisson structure 
can be recast into the form of a classical Yang-Baxter formalism.
Therefore the first step to consider the integrability of BSLE 
either in the Liouville sense or in the Lax sense is to find its 
Lax formalism, and to this end the famous prolongation approach 
(Walquist \& Estabrook (1975), Lu \& Li (1989a, 1989b)) is preferred.

To begin with, we introduce a transformation of independent variables

\begin{displaymath}
x_+ \rightarrow \frac {x_{0}+x_{1}}{2}, ~~~x_- \rightarrow 
\frac {x_{0}-x_{1}}{2},
\end{displaymath}

\noindent which leads to the changes $\partial_\pm \rightarrow 
\partial_0 \pm \partial_1$ of the derivatives.

Setting $\pi_{0}=\partial_{0}\phi$, $\pi_{1}=-\partial_{1}\phi$, the 
system (\ref{comp}) can be expressed by the following set of 
rank two differential forms on the space of variables 
$(x_0,~x_1,~\phi,~\psi_+,~\psi_-,~\pi_0,~\pi_1)$,

\begin{eqnarray}
& &\alpha_{1}=d\psi_{+}{\wedge}dx_{1}-dx_{0}{\wedge}d\psi_{+}
-3\psi_{-}e^{\phi}dx_{0}{\wedge}dx_{1}, \nonumber\\
& &\alpha_{2}=d\psi_{-}{\wedge}dx_{1}+dx_{0}{\wedge}d\psi_{-}
-3\psi_{+}e^{\phi}dx_{0}{\wedge}dx_{1}, \nonumber\\
& &\alpha_{3}=d\phi{\wedge}dx_{1}-\pi_{0}dx_{0}{\wedge}dx_{1}, \nonumber\\
& &\alpha_{4}=d\phi{\wedge}dx_{0}-\pi_{1}dx_{0}{\wedge}dx_{1}, \nonumber\\
& &\alpha_{5}=d\pi_{0}{\wedge}dx_{1}-d\pi_{1}{\wedge}dx_{0}
-(18\psi_{+}\psi_{-}e^{\phi}+4e^{2\phi})dx_{0}{\wedge}dx_{1}. \label{forms}
\end{eqnarray}

\noindent On the intersection with the space of independent variables 
$(x_0,~x_1)$ the system (\ref{comp}) will be reproduced. 
It is easy to check that the system (\ref{forms}) of two forms
generate a closed ideal in the sense that

\begin{displaymath}
d \alpha_i = \eta_{ij} \alpha_j
\end{displaymath}

\noindent for some one forms $\eta_{ij}$. Given the system (\ref{forms}),
we now assume that the (enlarged) prolongation (Lu \& Li (1989a, 1989b)) 
form takes the form

\begin{eqnarray}
\omega= - dT+F(\phi, \psi_{+}, \psi_{-}, \pi_{0}, \pi_{1})Tdx_{0}+
G(\phi, \psi_{+}, \psi_{-}, \pi_{0}, \pi_{1})Tdx_{1}, \label{prolong}
\end{eqnarray}

\noindent where $F$ and $G$ are functions of indicated variables 
taking values in some undetermined Lie algebra, and 
the newly introduced ``pseudopotential'' $T$ lies  in the group
generated by that Lie algebra.

From the integrability condition 

\begin{eqnarray*}
d\omega{\in}I(\omega, \alpha), 
\end{eqnarray*}

\noindent where $I(\omega, \alpha)$ is an ideal generated by the set 
$\{\alpha_{i}\}$ and $\{\omega\}$, we have the following equations for 
$F$ and $G$

\begin{eqnarray}
&F_{\psi_{+}}-G_{\psi_{+}}=0, &\nonumber\\ 
&F_{\psi_{-}}+G_{\psi_{-}}=0, &\nonumber\\
&F_{\pi_{0}}-G_{\pi_{1}}=0, &\nonumber\\
&F_{\pi_{1}}+G_{\pi_{0}}=0,& \nonumber\\
&{[F, G]+\pi_{1}{F_{\phi}}+\pi_{0}{G_{\phi}}
+3e^{\phi}(\psi_{-}{F_{\psi_{+}}}+
\psi_{+}{G_{\psi_{-}}})} & \nonumber\\
&-(18\psi_{+}{\psi_{-}}e^{\phi}+4e^{2\phi})F_{\pi_{1}}=0.& \label{eqns}
\end{eqnarray}

\noindent where $[F, G]=FG-GF$. Solving the system of equations (\ref{eqns}), 
we get 

\begin{eqnarray*}
& &F=-{\frac {1}{2}}[\pi_{1}L_{0}+3\psi_{+}e^{\frac {\phi}{2}}L_{1}-
3\psi_{-}e^{\frac {\phi}{2}}L_{-1}+e^{\phi}L_{2}-e^{\phi}L_{-2}], \nonumber\\
& &G={\frac {1}{2}}[\pi_{0}L_{0}+3\psi_{+}e^{\frac {\phi}{2}}L_{1}+
3\psi_{-}e^{\frac {\phi}{2}}L_{-1}+e^{\phi}L_{2}+e^{\phi}L_{-2}], 
\end{eqnarray*}

\noindent where $L_{i}$, $i=0, \pm1, \pm2$, are operators 
satisfying the following commutation relations:

\begin{equation}
\begin{array}{ll}
$$ [L_{0}, ~L_{1}]=-L_{1},$$ & $$ [L_{0}, ~L_{-1}]=L_{-1},$$ \cr  
$$ [L_{0}, ~L_{2}]=-2L_{2},$$ & $$ [L_{0}, ~L_{-2}]=2L_{-2}, $$\cr
$$ [L_{1}, ~L_{-1}]=2L_{0},$$ & $$ [L_{1}, ~L_{-2}]=3L_{-1}, $$ \cr
$$ [L_{-1}, ~L_{2}]=-3L_{1}, $$ & $$ [L_{2}, ~L_{-2}]=4L_{0}.$$ 
\end{array}
\label{gen}
\end{equation}

Notice that the system (\ref{gen}) does not yet generate a 
closed algebra. However, one can easily see that all relations in (\ref{gen})
can be rewritten in a unified form 

\begin{equation}
[L_n,~L_m] = (n-m) L_{n+m}, \label{witt}
\end{equation}

\noindent for $n,~m = 0,~\pm 1,~\pm 2$. Defining new generators 
iteratively by

\begin{eqnarray*}
L_{m+2}=\frac{1}{m} [L_{m+1}, ~L_{1}],~~~~ 
L_{-m-2}=\frac{1}{m} [L_{-1}, ~L_{-m-1}], ~~~~m\geq 1,
\end{eqnarray*}

\noindent then equation (\ref{witt}) will close
over the generators $L_{j}$, $j=0, \pm1, \pm2, ...$. This is the  
well-known Witt algebra or ``centerless Virasoro algebra''.

Now intersecting the prolongation form (\ref{prolong}) 
on the solution manifold $(x_+,~x_-)$, 
we obtain the Lax pair for BSLE (\ref{comp})

\begin{eqnarray}
& &\partial_{+}T=(F+G)T, \nonumber\\
& &\partial_{-}T=(F-G)T. \label{lax}
\end{eqnarray}

The existence of Lax pair (\ref{lax}) ensures that the BSLE (\ref{comp}) 
is integrable in the Lax sense. However since no Hamiltonian structure is
currently known for BSLE, the Liouville integrability cannot be established
at this point.

Notice that the Lax pair (\ref{lax}) involves the generator of Witt 
algebra with degrees ranging from $-2$ to 2. It is a well known fact that
the Witt algebra does not contain any finite dimensional subalgebra
of domension greater than 3. Therefore the Witt algebra is the only 
possible gauge algebra of the Lax system (\ref{lax}). Moreover, 
as there is no nondegenerate symmetric bilinear form on Witt algebra,
it is hard to obtain a Lagrangian formulation for BSLE as in the 
conventional Toda case by taking the trace of $A_+A_-$ with $A_\pm$ being 
the Lax potentials. Actually if the Lagrangian is indeed in the form 
of a trace over $A_+A_-$ in the case of BSLE, then it would lead to the 
conclusion that BSLE is a topological theory because the 
Lagrangian is identically zero. Whether this is true or not still 
deserves further study.

\section{SOLUTION OF BSLE}

Given the Lax pair (\ref{lax}), we can now consider the possible
solutions of the BSLE (\ref{comp}) using the Leznov-Saveliev analysis. 

For convenience we choose the following specific gauges for the Lax pair 
of BSLE,

\begin{eqnarray}
& &\partial_{+}T_{L}=(\partial_{+}{\phi}L_{0}+3\psi_{+}L_{1}+L_{2})T_{L}, 
\nonumber\\
& &\partial_{-}T_{L}=-(3\psi_{-}e^{\phi}L_{-1}+e^{2\phi}L_{-2})T_{L}, 
\label{laxl}
\end{eqnarray}

\noindent and

\begin{eqnarray}
& &\partial_{+}T_{R}=(3\psi_{+}e^{\phi}L_{1}+e^{2\phi}L_{2})T_{R}, \nonumber\\
& &\partial_{-}T_{R}=-(\partial_{-}{\phi}L_{0}+3\psi_{-}L_{-1}+L_{-2})T_{R}, 
\label{laxr}
\end{eqnarray}

\noindent where 

\begin{eqnarray}
T_{L}=e^{{\frac {\phi}{2}}L_{0}}T,~~~~ 
T_{R}=e^{-{\frac {1}{2}}{\phi}L_{0}}T.
\end{eqnarray}

Now let us choose some highest weight representation of the Witt algebra
with highest weight $h$ and denote the highest weight vector by $|h \rangle$.
The dual of $| h \rangle$ is denoted $\langle h |$. The highest weight 
conditions read

\begin{eqnarray}
& &L_{0} |h\rangle= h|h\rangle, ~~~~\langle h|L_{0}=\langle h| h, \nonumber\\
& &L_{n}|h\rangle=0, ~~~~\langle h|L_{-n}=0,~~~~ (n>0), \nonumber\\
& &\langle h|h\rangle=1.  \label{highest}
\end{eqnarray}

From (\ref{laxl}), (\ref{laxr}) and (\ref{highest}), it follows that 

\begin{eqnarray}
{\langle}h|\partial_{-}T_{L}=0, ~~~~~\partial_{+}T_{R}^{-1}|h\rangle=0,
\label{tlr}
\end{eqnarray}

\noindent and hence the vectors 

\begin{eqnarray}
\xi(x_{+})={\langle}h|T_{L}, ~~~~~\bar{\xi}(x_{-})=T_{R}^{-1}|h\rangle, 
\end{eqnarray}

\noindent are chiral, namely

\begin{eqnarray}
\partial_{-}\xi(x_{+})=0, ~~~~~\partial_{+}\bar{\xi}(x_{-})=0.
\end{eqnarray}

\noindent Moreover, defining 

\begin{eqnarray}
& &{\bar{T}}_{L}=e^{\psi_{+}L_{-1}}T_{L}, \nonumber\\
& &{\bar{T}}_{R}=e^{-\psi_{-}L_{1}}T_{R}, \nonumber
\end{eqnarray}

\noindent an easy calculation leads to

\begin{eqnarray}
{\langle}h|L_{1}\partial_{-}{\bar{T}}_{L}=0, ~~~~~
\partial_{+}{\bar{T}_{R}}^{-1}L_{-1}|h\rangle=0, 
\end{eqnarray}

\noindent showing that the vectors 

\begin{eqnarray}
\zeta(x_{+})={\langle}h|L_{1}{\bar{T}}_{L},~~~~~ \bar{\zeta}(x_{-})
=\bar{T}_{R}^{-1}L_{-1}|h\rangle, 
\end{eqnarray}

\noindent also are chiral

\begin{eqnarray}
\partial_{-}\zeta(x_{+})=0, ~~~~\partial_{+}\bar{\zeta}(x_{-})=0.
\label{zeta}
\end{eqnarray}

From equations (\ref{tlr}-\ref{zeta}), a straightforward calculation gives

\begin{eqnarray}
& &\xi(x_{+}){\bar{\xi}} (x_{-})=e^{h\phi}, \nonumber\\
& &\zeta(x_{+}){\bar{\xi}}(x_{-})=2h\psi_{+}e^{h\phi}, \nonumber\\
& &\xi(x_{+}){\bar{\zeta}}(x_{-})=2h\psi_{-}e^{h\phi}, \nonumber
\end{eqnarray}

\noindent which in turn, gives a formal solution to BSLE 

\begin{eqnarray}
& &\phi={\frac {1}{h}}{\rm ln} (\xi(x_{+})\bar{\xi}(x_{-}),\\ 
& &\psi_{+}={\frac {1}{2h}}{\frac {\zeta(x_{+})
\bar{\xi}(x_{-})}{\xi(x_{+})\bar{\xi}(x_{-})}}, \\
& &\psi_{-}= {\frac {1}{2h}}{\frac {\xi(x_{+})\bar{\zeta}(x_{-})}
{\xi(x_{+})\bar{\xi}(x_{-})}}. \label{sol}
\end{eqnarray}

Some remarks are in due course. First, one could be quite dubious on the 
correctness of the assumption of the highest weight conditions 
(\ref{highest}). Indeed, it is known from the study of comformal field 
theory that no nontrivial {\it unitary} highest weight representations
exists for Virasoro algebra at center $c=0$. However, as we are using 
the Witt algebra as a gauge algebra of our Lax system, wo do not 
concern the unitarity of the representation and so are free to choose the 
non-unitary representations in (\ref{highest}). Actually, the 
choice of non-unitary representations in (\ref{highest}) is not
unavoidable if we introduce an extra auxiliary field, say $\rho$, and modify 
the Lax system (\ref{lax}) to the form

\begin{eqnarray}
& &\partial_{+}T=(\partial_+ \rho c+ F+G)T, \nonumber\\
& &\partial_{-}T=(- \partial_- \rho c + F-G)T,  \label{laxmod}
\end{eqnarray}

\noindent and, in the mean time, change the gauge algebra (\ref{witt})
into the full Virasoro algebra

\begin{displaymath}
[L_n,~L_m] = (n-m) L_{n+m} + \frac{c}{12}(n^3-n) \delta_{n+m,~0}. 
\end{displaymath}

\noindent One can show that such modifications do not change the equations
of motion for $\phi,~\psi_\pm$ and only give rise to a new equation for 
the auxiliary field $\rho$,

\begin{displaymath}
\partial_+ \partial_- \rho + 2 {\rm exp}(2 \phi)=0 .
\end{displaymath}

\noindent The modified Lax system (\ref{laxmod}) can then be treated
in exactly the same way as above and one can choose unitary highest weight 
representations of the Virasoro algebra in place of the non-unitary 
representations in (\ref{highest}).

Another remark is as follows. Though the solution of BSLE (\ref{comp})
can be expressed in the form of (\ref{sol}), the chiral vectors
cannot be regarded as arbitrary because they are defined from the 
non-chiral objects $T_L,~T_R$ and $\bar{T}_L,~\bar{T}_R$ subjecting 
nontrivial constraints (the Lax pair). The explicit solution of BSLE
therefore connot be obtained in this way. In conventional Liouville and Toda
cases, one can, however, make a similar construction starting not from the 
specific gauges (\ref{laxl}) and (\ref{laxr}) of the Lax pair but from the
set of so-called Drinfeld-Sokolov systems. In the present case
such systems would look like

\begin{eqnarray}
& &\partial_+ Q = (\partial_+ k(x_+) L_0 + 3 p(x_+) L_1 + L_2)Q,~~~
\partial_- Q=0,\nonumber\\
& &\partial_- \bar{Q} =  \bar{Q} (\partial_- k(x_-) L_0 + 3 p(x_-) L_{-1} 
+ L_{-2}),~~~\partial_+ \bar{Q}=0 \nonumber
\end{eqnarray}

\noindent with some {\it arbitrary} chiral functions $k(x_\pm)$ and 
$p(x_\pm)$. Unfortunately we have been unable to obtain exact solutions to
(\ref{comp}) using the above Drinfeld-Sokolov systems.

\section{DISCUSSION}

In this paper, we have identified the Lax integrability for 
BSLE (\ref{comp}) using the enlarged prolongation approach. 
The same set of Lax pair will be obtained if we used the original 
scalar form of prolongation forms as did by Estabrook and Walquist
in their classical paper Walquist \& Estabrook (1975).                      

On the other hand we expressed the solution of BSLE in terms of some 
chiral vectors obtained from the action of solution of the Lax system 
in some specific gauges which is in complete analogy to the 
conventional Toda and bosonic Toda cases. 
However, as is mentioned in the end of the last section, the 
Drinfeld-Sokolov construction of solutions for BSLE is not 
established and this may be one of the subtle points where the BSLE 
behaves different from the bosonic super Toda theories.

To conclude this paper, we briefly point out some other related open 
problems:

(1) Its easy to see that, the quantity ${\frac {1}{2}}\partial_{+}
\phi\partial_{-}\phi+18\psi_{+}\psi_{-}e^{\phi}+2e^{2\phi}$ is a 
Lagrangian for the ``Liouville part'' $\phi$ in BSLE (\ref{comp}).
However no Lagrangian expression for $\psi_\pm$ is currently known.
A principal reason is that $\psi_{\pm}$ are chiral fields of first order, 
and it seems interesting to see whether one can construct a Hamiltonian 
or Lagrangian formalism for the whole BSLE via the Dirac method.

(2) It was shown in the introduction that the BSLE 
is conformal invariant and thus admit a $Witt_L \otimes Witt_R$ symmetry 
algebra. Is there any relationship between the conformal symmetry
algebra and the gauged Witt algebra?

\newpage

\section*{REFERENCES}

\noindent Au, G., Spence, B., (1995). Modern Physics Letters {\bf A10}, 2157.

\noindent Chao, L., (1993). Commmunications in Theoretical Physics 
{\bf 20}, 221.

\noindent Chao, L., (1994) {\it Toda fields in $d\geq2$-dimensions}, 
Talk Given at the International Workshop on the Frontiers of Quantum 
Field Theory, ITP Academy of China, Beijing, China.

\noindent Chao, L., (1995). preprint hep-th/9512198.

\noindent Chao, L., Hou, B.Y., (1993). International Journal of Modern Physics 
{\bf A8}, 3773.

\noindent Chao, L., Hou, B.Y., (1994). Annual of Physics (New York) {\bf 230}, 1.

\noindent Hou, B.Y., Chao, L., (1993). Int. J. Mod. Phys. {\bf A8}, 1105.

\noindent Ferreira, L.A., Gervais, J.-L., Sanchez Guillen, J., Saveliev, M.V.,
(1995). Preprint hep-th/9512105.

\noindent Gervais, J.-L., Saveliev, M.V., (1995). Preprint hep-th/9505047.

\noindent Lu, J.F., Li, Y.Q., (1989a). Physics Letters {\bf A135}, 179.

\noindent Lu, J.F., Li, Y.Q., (1989b). Comm. Theore. Phys. {\bf 11}, 99.

\noindent Toppan, F. (1991). Physics Letters {\bf B260}, 346.

\noindent Wahlquist, H., Estabrook, F.B. (1975). 
Journal of Mathematical Physics {\bf 16}, 1.

\end{document}